\documentclass[12pt]{article}

\begin{document}
\title{Bell inequalities and \\ incompatible measurements}
\author{Peter Morgan\\
     \small 30, Shelley Road, Oxford, OX4 3EB, England.\footnote{Address until August 2003:
   207, von Neumann Drive, Princeton NJ08540, USA.}\\
     \small peter.morgan@philosophy.oxford.ac.uk\\
     \small http://users.ox.ac.uk/{\footnotesize$\!\sim$}sfop0045}
\date{\today}
\maketitle

\begin{abstract}
Bell inequalities are a consequence of measurement incompatibility (not, as generally
thought, of nonlocality).
In classical terms, this is equivalent to contextuality --- measurement devices do
have a significant effect.
Contextual models are reasonable in classical physics, which always took the view that
we ignore measurement devices whenever possible, but if that isn't good enough 
then we do have to model measurement devices.
It is also argued that quantum theory should only be taken with counterfactual
seriousness, because measurement incompatibility is a counterfactual concept.
\end{abstract}
\maketitle

\setlength{\parskip}{6pt}
\section{introduction}
The first part of this paper, section \ref{BellIneq}, addresses the violation of Bell
inequalities, and shows that a classical contextual way of thinking about the experiments
is quite possible.

The second part, section \ref{IncompMeas}, shows that more generally than for experiments
that violate Bell inequalities, the incompatibility of measurements is a result of a classically
remarkable construction: even if an empirically adequate classical noncontextual model of a
measured system is not possible, we nonetheless describe the measured system and a
number of measurement systems separately, using a quantum state and operators on a
Hilbert space to represent them.

Section \ref{IncompMeas} also argues that the idea of ``incompatible measurements''
depends on a counterfactual relocation of measurements that are in fact compatible.

\section{Bell inequalities}\label{BellIneq}
The data that comes from an experiment that violates Bell inequalities can be summarized
in a table with four columns, two for each arm of the experiment, ``$A$'' and ``$a$'' for
one arm and ``$B$'' and ``$b$'' for the other arm (see table \ref{RawData}).
Each $A$ is an actual measurement result, and each $a$ is an actual measurement setting,
and similarly for $B$ and $b$.
From these experimental results we can construct statistics, but we'll do the usual thing
of assuming that we've taken so many measurements that the statistics can be thought of in
an idealized way as probabilities. The experimental data can be summarized as a classical
probability distribution $p(A,a,B,b)$.

A little closer to experiment would be two tables of data, ``$A$'', ``$a$'', and
``$t_A$'' for one arm (where $t_A$ is the time at which we observe $A$ and $a$) and
``$B$'', ``$b$'', and ``$t_B$'' for the other arm (see table \ref{RawData}).
To construct a table with just $A, a, B,$ and $b$, we look for times $t_A$ and $t_B$ that
match within some tolerance, and put the corresponding values of $A, a, B,$ and $b$ into
our summary table.

\begin{table}[htb]
\centerline{\begin{tabular}{c c @{\hspace{4em}} c}
      \multicolumn{2}{c}{} &
     {Selected by} \\
      \multicolumn{2}{c}{Raw data:} &
     {$t_A\approx t_B$} \\
      \multicolumn{2}{c}{} &
     {$[{\scriptstyle p(A,a,B,b)}]$} \\
\begin{tabular}{| c | c | c |}\hline $A$ & $a$ & $t_A$ \\ \hline
  $\vdots$ & \begin{tabular}{c} $a_1$\\ $a_2$\\ $a_1$\\ $a_1$\\ $a_2$\\
  $\vdots$\end{tabular} &
  $\vdots$ \\ \hline
\end{tabular}&
\begin{tabular}{| c | c | c |}\hline $B$ & $b$ & $t_B$ \\ \hline
  $\vdots$ & \begin{tabular}{c} $b_1$\\ $b_1$\\ $b_2$\\ $b_2$\\ $b_2$\\
  $\vdots$\end{tabular} &
  $\vdots$ \\ \hline
\end{tabular}&
\begin{tabular}{| c | c | c | c |}\hline $A$ & $a$ & $B$ & $b$ \\ \hline
  $\vdots$ & \begin{tabular}{c} $a_1$\\ $a_2$\\ $a_1$\\ $a_1$\\ $a_2$\\
  $\vdots$\end{tabular} &
  $\vdots$ & \begin{tabular}{c} $b_1$\\ $b_1$\\ $b_2$\\ $b_2$\\ $b_2$\\
  $\vdots$\end{tabular} \\ \hline
\end{tabular}
\end{tabular}}
\caption{\label{RawData} The raw experimental data, and the data selected \newline
\hspace*{5em} by approximately matching times of measurement.}
\end{table}

To keep things simple, we'll have to idealize again: $a$ can take one of only two values,
$a_1$ or $a_2$; and $b$ can take one of only two values, $b_1$ or $b_2$.
From the table of $A$, $a$, $B$, and $b$, which is a completely classical description of
the experimental results, we can select data that corresponds to each of four possible
cases, $(a=a_1,b=b_1)$, $(a=a_1,b=b_2)$, $(a=a_2,b=b_1)$, and $(a=a_2,b=b_2)$, giving us
four tables of actual data (see table \ref{FourTables}), which in an idealized description
can be represented by four probability distributions,
$p(A_1,B_1)$, $p(A_1,B_2)$, $p(A_2,B_1)$, and $p(A_2,B_2)$.

\begin{table}[htb]
\centerline{\begin{tabular}{c c c c}
     {Selected by} & {Selected by} & {Selected by} & {Selected by} \\
     {$t_A\approx t_B$} & {$t_A\approx t_B$} &
     {$t_A\approx t_B$} & {$t_A\approx t_B$} \\
     {$a=a_1$} & {$a=a_1$} & {$a=a_2$} & {$a=a_2$} \\
     {$b=b_1$} & {$b=b_2$} & {$b=b_1$} & {$b=b_2$} \\
     {$[{\scriptstyle p(A_1,B_1)}]$} & {$[{\scriptstyle p(A_1,B_2)}]$} &
     {$[{\scriptstyle p(A_2,B_1)}]$} & {$[{\scriptstyle p(A_2,B_2)}]$} \\
\begin{tabular}{| c | c |}\hline $A_1$ & $B_1$ \\ \hline
  \begin{tabular}{c} {} \\ $\vdots$ \\ {}\end{tabular} &
  \begin{tabular}{c} {} \\ $\vdots$ \\ {}\end{tabular}\\ \hline
\end{tabular}&
\begin{tabular}{| c | c |}\hline $A_1$ & $B_2$ \\ \hline
  \begin{tabular}{c} {} \\ $\vdots$ \\ {}\end{tabular} &
  \begin{tabular}{c} {} \\ $\vdots$ \\ {}\end{tabular}\\ \hline
\end{tabular}&
\begin{tabular}{| c | c |}\hline $A_2$ & $B_1$ \\ \hline
  \begin{tabular}{c} {} \\ $\vdots$ \\ {}\end{tabular} &
  \begin{tabular}{c} {} \\ $\vdots$ \\ {}\end{tabular}\\ \hline
\end{tabular}&
\begin{tabular}{| c | c |}\hline $A_2$ & $B_2$ \\ \hline
  \begin{tabular}{c} {} \\ $\vdots$ \\ {}\end{tabular} &
  \begin{tabular}{c} {} \\ $\vdots$ \\ {}\end{tabular}\\ \hline
\end{tabular}
\end{tabular}}
\caption{\label{FourTables} The data selected according to measurement settings.}
\end{table}

Now it's a simple fact\cite{deMuynck} that we can't always construct a probability
distribution $p(A_1,A_2,B_1,B_2)$ that has $p(A_1,B_1)$, $p(A_1,B_2)$, $p(A_2,B_1)$, and
$p(A_2,B_2)$ as marginal probability distributions, when these have been constructed
by selection from our table of values of $A$, $a$, $B$, and $b$.
This is often held to make a classical description of the world impossible.

More careful accounts don't say that.
Here, it's obvious that we've got a perfectly good classical description, $p(A,a,B,b)$,
and the fact that we can't describe everything using a probability distribution like
$p(A_1,A_2,B_1,B_2)$ may be awkward, but it's not impossible.
The technical difference between these two probability distributions is that
$p(A_1,A_2,B_1,B_2)$ is a ``non-contextual'' description, whereas $p(A,a,B,b)$ is a
``contextual'' description, because by including ``$a$'' and ``$b$'' we include information
about the measurement apparatus.
Modern Physics usually insists that a contextual description is not OK, but all we have
to do to let us think classically is to make sure that any description of a Bell experiment
is constructed to have $p(A,a,B,b)$ as a marginal probability distribution, not
$p(A_1,A_2,B_1,B_2)$ (which cannot be empirically adequate).
In other words, we have to remember that the experimental apparatus will often have to
be part of the model.
This shouldn't worry us, because classical physics always took the view that we
ignore the experimental apparatus if we can, but if it isn't good enough to do that then we
\emph{do} also have to model the experimental apparatus.

What about \emph{nonlocality}?
Despite everything that's been written about nonlocality, it's largely irrelevant: it's
contextuality that's the main issue.
It's an experimental fact so far that we can't send messages faster than light:
$p(A,a,B,b)$, in particular, is part of that experimental fact (satisfying $p(A|a,b)=p(A|a)$
and $p(B|a,b)=p(B|b)$, which together ensure the necessary statistical independence).
A classical description of the world doesn't have to send messages faster than light, and
mustn't allow messages to be sent faster than light, in models of existing experiments.

Nonetheless, there is a classical point of view from which the experimental data
$p(A,a,B,b)$ does look strange.
If we think there are classical \emph{particles}, for which there is an essentially
unavoidable idea that they are separate from one another, we will tend to get
ourselves into the difficulties that are usually associated with Bell inequalities
(although there are what may be called exotic loopholes).
If we instead think in terms of classical \emph{fields}, and make sure there is no idea
of separable particles, we won't get ourselves in those difficulties (in particular,
see ``The derivation of Bell inequalities for beables''\cite{MorganBell}).
A classical theory will have to be a classical statistical theory,
and for a classical statistical theory to be good enough, it will have to be a classical
statistical field theory, just as for a quantum theory to be good enough it has to be a
quantum field theory.
Quantum theory has a great advantage here, because there are many interesting
experiments where a quantum field theory model can be reduced quite effectively
to a model that uses two or three quantum particles, or even a finite dimensional
Hilbert space, whereas for the same situation a classical statistical field theory model
can't be reduced to as simple a model in terms of classical particles --- wave dynamics
and interactions with measurement devices remain important.

\section{Incompatible measurements}\label{IncompMeas}
Two assumptions are generally made in physical experiments.
Firstly, successive runs of an experimental apparatus are separated by sufficient time
for the results to be statistically independent.
Secondly, if several apparatuses are used, they are separated by sufficient distance
for the results to be statistically independent.
It is generally held to be a requirement of any physical theory that the second
assumption is satisfiable, which is codified for quantum field theory as the
cluster decomposition principle\cite{Weinberg}.
The first assumption has been taken to be problematic in the foundations of
classical statistical mechanics, where it has been an aim to discover conditions on
which it can be proved that an ensemble of a single system considered at different
times is equivalent to an ensemble of many systems in different places.
In practice, we seek to ensure that a number of runs of a single experimental apparatus
are empirically equivalent to having available to us a large number of experimental
apparatuses at quite large space-like separations from each other.

Assume here, therefore, that we carry out an experiment using a set
of experimental apparatuses at quite large space-like separations from each other.
The measurements we make of these experimental apparatuses can be tabulated (just as
we tabulated the results of an experiment that violates Bell inequalities in section
\ref{BellIneq}), and we can again select results from the table conditional on the
measurement settings to give many tables, which we take to be represented in quantum
theory by noncommutative operators.

From a quantum field theoretical point of view, however, in a description of the many
experimental apparatuses that is as close as possible to experimental detail, all
the measurements correspond to mutually compatible operators because they are
measurements associated with regions of space-time that are at space-like
separation.
The concept of incompatible observables depends on the idea that actually measured
compatible observables associated with space-like separated regions (or with time-like
separated regions that are far enough apart) would not have been compatible if they
had been measured in the same region.
The measurements are incompatible only if we take them to have been carried out
in a single space-time region, when in fact they clearly have not been.
Although such a concept is mathematically very useful, we do not have to take it to be
fundamental.

Quantum field theory is more empirically adequate than quantum particle theory, and we
usually agree that a quantum field theory reduces to a quantum particle theory in a 
suitable approximation, so our discussion will be framed here in terms of quantum field theory.
A quantum field theoretical description of the whole set of experimental apparatuses
is very underdetermined by the measurements we make, simply because they are all
compatible measurements, as well as because we make only a finite number of measurements.
The set of candidate quantum field states that describe the whole set of experimental
apparatuses in fact must include quantum field states that have a representation as a
non-negative Wigner function (again just because the measurements we make are mutually
compatible; examples of Wigner functions for quantum field states can be
found in ``A relativistic variant of the Wigner function''\cite{MorganWigner}).
A representation of a quantum field state as a non-negative Wigner function can be
regarded as a classical state. Consequently, a classical contextual description of
any experiment is entirely possible (that is, an empirically adequate classical
description of the measurement systems and the measured systems together is possible
whenever an empirically adequate quantum field description of the measurement systems
and the measured systems together is possible).

To obtain a quantum field state that is acted on by a number of incompatible observables
as an alternative description of the whole set of experimental apparatus, we have to identify
a class of measurements that will be termed ``measurement settings'' and a distinct class
of measurements that will be termed ``measurement results'', as we did in section
\ref{BellIneq}, to allow us to construct a table of measurement results for each measurement
setting.
Although this separation is possible and very useful, it is also arbitrary, an arbitrariness
corresponding to the notorious ``Heisenberg cut'', and in general the separation ignores
the classical impossibility of a noncontextual model.

A detailed description of a measured system without considering its interactions with
a measurement system is not generally possible in classical physics.
We are of course very happy when a description of a measured system alone is adequate,
since reduced complexity is welcome, but classical physics never required that this
must be possible.
We can present the possibility of classical description without considering the
whole of a measurement apparatus as the possibility of constructing an empirically
adequate quantum state that has a representation as a non-negative Wigner function.

In quantum theory, we seek to describe the interaction between a single preparation
device and many different measurement devices in a unified way.
\begin{list}{$\bullet$}{}
\item
If we think classically and noncontextually, the preparation device prepares the same
classical field (or particle) state independently of what measurement device is used.
This is generally not empirically adequate.
\item
If we think quantum theoretically, the preparation device prepares the same quantum
state independently of what measurement device is used, even if an empirically
adequate classical noncontextual model is not possible.
This has been found to be empirically more adequate --- but we do have to consider the
measurement device explicitly if we want very great detail (for example, it may be
significant that the measurement device is not at absolute zero temperature; all non-ideal
properties will have to be considered eventually).
\item
If we think classically and contextually, the preparation device does not prepare
the same classical field state independently of what measurement device is used.
Although it is a shame to lose the effectiveness of quantum theory for cases where we can
adequately think of a quantum state being independent of measurement devices, it may
also be theoretically preferable to represent all the interactions of the preparation
device with the many different measurement devices in a unified way.
\end{list}

From the Copenhagen interpretation's point of view, all description of experiments is
given in classical terms.
We can only follow the Copenhagen interpretation in its construction of a quantum
state, however, if we make a distinction between measurement settings and
measurement results that is in general not classically motivated.
Although we can usefully make this distinction, other approaches may also be
useful, and more detailed classical description of experiments is one possibility.

\section{Conclusion}
Although quantum theory is very useful, it rests on a counterfactual assumption:
we can carry out incompatible measurements.
Although it is very effective to counterfactually relocate actual, compatible 
measurements (or very nearly compatible measurements if they are at large
time-like separation) to a common region of space-time, where they are, from
the point of view of quantum theory, incompatible, we are not justified in taking
quantum theory with more than counterfactual seriousness.

Also problematic is the separation of measurement devices from measured systems,
which quantum theory can do in its terms, even when classical separation is not
possible. Although this is very effective when it can be done, it may be theoretically
preferable to represent all the interactions of measurement devices with
measured systems in a unified way.

\end{document}